\documentclass[%
 reprint,
superscriptaddress,
amsmath,amssymb,
aps,
floatfix,
]{revtex4-2}

\usepackage{graphicx}% Include figure files
\usepackage{dcolumn}% Align table columns on decimal point
\usepackage{bm}% bold math
\usepackage[colorlinks,urlcolor=blue,citecolor=blue,linkcolor=blue]{hyperref}
\usepackage{svg}
\usepackage{braket}
\usepackage{float}
\usepackage{hyperref}

\newcommand{\fprime}{F^{\prime}}
\begin{document}
\preprint{APS/123-QED}

\title{Controlling light shifts in chip-scale atomic beam clocks}
\author{Alexander~Staron}
\email{Contact author: alexander.staron@colorado.edu}
\affiliation{Time and Frequency Division, National Institute of Standards and Technology, Boulder, CO, USA}
\affiliation{Department of Physics, University of Colorado Boulder, Boulder, CO, USA}

\author{Mingwu~Lu}
\altaffiliation{Present address: Infleqtion, Boulder, CO, USA}
\affiliation{Time and Frequency Division, National Institute of Standards and Technology, Boulder, CO, USA}
\affiliation{Department of Physics, University of Colorado Boulder, Boulder, CO, USA}

\author{Ruwan~Senaratne}
\affiliation{HRL Laboratories, Malibu, CA, USA}

\author{Travis~Autry}
\altaffiliation{Present address: DARPA, Arlington, VA, USA}
\affiliation{HRL Laboratories, Malibu, CA, USA}

\author{Susan~Schima}
\author{John~Kitching}
\author{William~McGehee}
\affiliation{Time and Frequency Division, National Institute of Standards and Technology, Boulder, CO, USA}

\date{\today}% It is always \today, today,

\begin{abstract}
Chip-scale atomic beam clocks are being investigated to extend the range of clock stability achievable in low-power timing applications. Here, we demonstrate a centimeter-scale, Ramsey coherent population trapping (CPT) clock based on a microfabricated Cs atomic beam device and investigate the interplay between light shifts and Doppler shifts that determines its leading clock systematics. We show that these shifts exhibit competing dependencies on CPT light parameters, leading to ``doubly-insensitive" operating points where the clock frequency is simultaneously insensitive to laser frequency and power. We further demonstrate a method for controlling key clock shifts using spectroscopic signatures from the atomic beam that is compatible with fully-integrated operation. The clock achieves a fractional frequency stability of $2 \times 10^{-10}$ at $1~\textrm{s}$ and sub-$\mu$s drift over nearly $17~\textrm{hours}$, with leading CPT light systematics controlled below the $10^{-12}$ level.
\end{abstract}

\maketitle

\section{Introduction}
\label{sec:intro}

Stable atomic timing with low size, weight, and power consumption (SWaP) is critical for a wide range of modern technologies. Chip-scale atomic clocks (CSACs) based on microwave, coherent population trapping (CPT) resonances are the leading low-power atomic clock technology and currently provide $\mu\textrm{s}$-level timing over approximately one day~\cite{vanier_atomic_2005, kitching_chip-scale_2018}. At longer timescales, however, CSACs exhibit frequency drift due to light shifts, collisional shifts, and cell aging. Methods to reduce long-term systematics include advanced interrogation protocols and interleaved measurement techniques for suppressing light shifts~\cite{yudin_combined_2018, shuker_reduction_2019, shuker_ramsey_2019, yudin_general_2020, zhu_theoretical_2000, shah_continuous_2006, mejri_atomic_2016, zhang_rubidium_2016, yudin_generalized_2018, abdel_hafiz_symmetric_2018, carle_pulsed-cpt_2023}, as well as improved cell fabrication strategies that limit gas permeation and aging effects~\cite{dellis_low_2016, carle_reduction_2023, carle_reduction_2024}. These approaches have not yet been widely realized in fully integrated, low-SWaP devices.

Chip-scale atomic beam clocks (CSABCs), first demonstrated in~\cite{martinez_chip-scale_2023}, offer a promising alternative to vapor-cell CSACs for low-SWaP, low-drift timing. Directed, buffer-gas-free beam propagation mitigates alkali–alkali collisional shifts~\cite{micalizio_spin-exchange_2006} and eliminates buffer gas-related frequency shifts~\cite{kozlova_temperature_2011}, removing two potential sources of long-term drift. In addition, Ramsey–CPT interrogation can reduce sensitivity to light shifts relative to continuous interrogation~\cite{micalizio_metrological_2012, yano_theoretical_2014}. As a result, CSABCs provide a distinct systematic landscape that may enable improved long-term stability while remaining compatible with chip-scale integration.

Despite these advantages, light shifts remain a dominant systematic in CSABCs. Frequency shifts can be induced by variation in essentially all laser properties (laser detuning, optical power, sideband modulation index, polarization, and beam pointing), and the magnitude of these shifts can vary dramatically with the interrogation scheme chosen. In previous work with a Rb-based CSABC, we identified a competition between resonant light shifts~\cite{hemmer_ac_1989, shahriar_darkstate_1997,vanier_atomic_2005} and Doppler shifts~\cite{mungall_cavity_1972, lee_accuracy_1995,esnault_cold-atom_2013} that suppressed the sensitivity to laser frequency fluctuations by more than two orders of magnitude at specific operating points~\cite{staron_doppler-shift_2026}. This cancellation arises from asymmetric decay in the CPT $\Lambda$-system and variation in the CPT pumping rate across the measured atoms, motivating further study of mitigation strategies for the remaining light-based shifts that leverage the unique physics of the CSABC system.

\begin{figure*}[htbp]
\includegraphics[width=\textwidth]{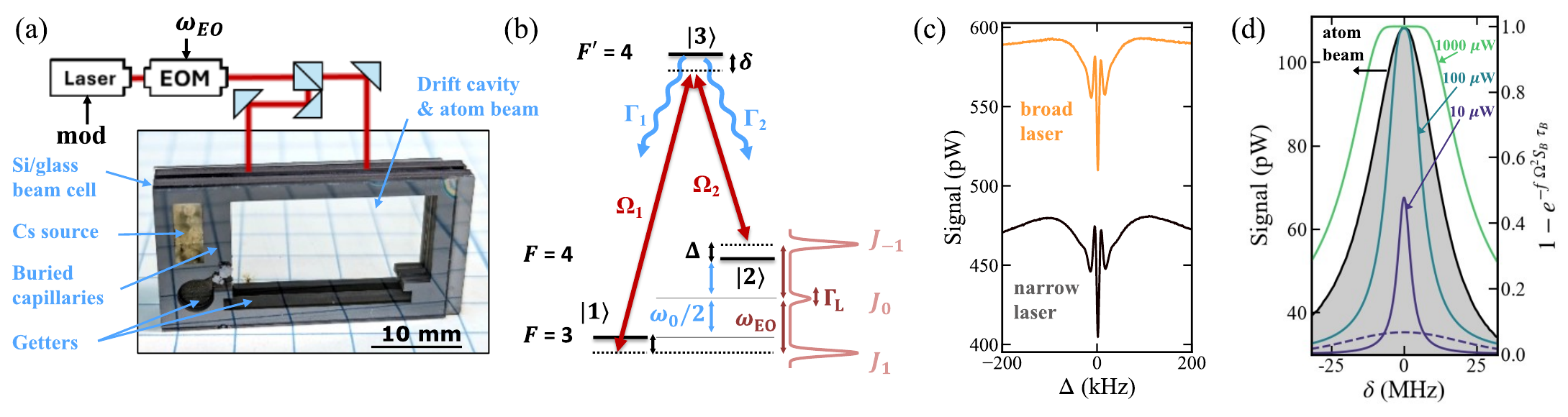}
\caption{Ramsey CPT spectroscopy using a Cs chip-scale atomic beam. (a) Image of the atomic beam device and laser schematic. (b) Energy level diagram of the CPT $\Lambda$-system showing relevant CPT Rabi rates ($\Omega_1$ and $\Omega_2$ in red), decay rates ($\Gamma_1$ and $\Gamma_2$ in blue), detunings ($\delta$ and $\Delta$), and laser spectrum (light red). (c) Ramsey CPT fluorescence spectra are shown using narrow ($\Gamma_{\rm L} \ll \Gamma$, black) and broad ($\Gamma_{\rm L} \approx 9.3 \ \Gamma$, orange) laser linewidths. (d) Measured atomic fluorescence versus laser detuning (black) at low probe power ($\approx 7~\mu\textrm{W}$) shows the beam's transverse Doppler detuning distribution. The CPT readout probability (Eq.~\ref{eq:rho33}) varies across the atomic beam and is plotted for a narrow laser at $\delta = 0$ at three different total optical powers (purple, blue, and green solid lines). The dashed line indicates the readout probability for a spectrally broad laser with $\Gamma_{\rm L} \approx 9.3 \ \Gamma$ at $P_{\rm total} \approx 10~\mu\textrm{W}$.}
\label{fig1}
\end{figure*}

Here, we demonstrate a comprehensive approach for managing light shifts in a new, Cs-based CSABC. We expand on our previously developed Doppler-cancellation strategy, showing that nulled laser detuning sensitivity is achievable in Cs and that ``doubly insensitive" operation points with simultaneously nulled laser-power and laser-detuning sensitivity exist. We find that realizing these points is largely controlled by modifications to the Doppler shift set by the laser linewidth and the atomic detuning distribution. The off-resonant light shifts are isolated and the interplay with resonant shifts is measured to explore the modulation index sensitivity. We demonstrate decoupling of these two shifts and explore both passive and active control of the modulation index. Together, these strategies suppress key light-based systematics without increasing device complexity, and we demonstrate control of each of the light-based systematics below the $10^{-12}$ level beyond $10^{4}~\textrm{s}$.

\section{Ramsey-CPT Spectroscopy on a cm-Scale Cs Beam}
\label{sec:methods}

We realize a CSABC using Ramsey–CPT spectroscopy of thermal Cs atoms in a prototype chip-scale atomic beam device, similar to Refs.~\cite{martinez_chip-scale_2023, staron_doppler-shift_2026}. The device, shown in Fig.~\ref{fig1}(a), is fabricated from five alternating layers of Si and glass that are anodically bonded to form a hermetic vacuum package with a total volume of $\sim 3.5~\textrm{mL}$. The device implements an effusive beam architecture in which alkali vapor generated in a smaller ``source” cavity passes through a buried microcapillary array into a larger ``drift” cavity, where spectroscopy is performed. The microcapillary array~\cite{li_cascaded_2019, li_robust_2020} is etched into one of the Si layers near the vertical center of the device and consists of ten parallel channels, each with length $3~\textrm{mm}$ and a square cross section of $0.1~\textrm{mm} \times 0.1~\textrm{mm}$. This channel geometry defines the atomic beam collimation and thereby sets the angular distribution of the atomic ensemble.

The atomic beams propagate freely across the drift region, where the vacuum is estimated to be below $ 1~\textrm{Pa}$~\cite{martinez_chip-scale_2023}. The vacuum is maintained using non-evaporable getters, while background alkali vapor is strongly gettered by graphite rods to sustain a large differential alkali pressure between the source and drift cavities. For typical operation, the device is heated to $\approx 360~\textrm{K}$, producing an estimated atomic flux of $\approx 5 \times 10^{11}~\textrm{s}^{-1}$.

% Ramsey-CPT spectroscopy (transition & laser frequency stabilization):
Clock spectroscopy is performed by interrogating the Cs ground-state hyperfine splitting ($\omega_0 = 2\pi \times 9.192631770~\textrm{GHz}$) through optical excitation of the $D_1$ line (wavelength $\approx 894.5~\textrm{nm}$, decay rate $\Gamma = \Gamma_1 + \Gamma_2 \approx 2\pi \times 4.56~\textrm{MHz}$), as shown in Fig.~\ref{fig1}(b). Optical fields are generated using a volume Bragg grating external-cavity laser with a servoed linewidth $\Gamma_L < 2\pi \times 100~\textrm{kHz}$. A fiber-based electro-optic modulator (EOM) driven at $\omega_{\rm EO} \approx \omega_0/2$ generates first-order sidebands that drive the CPT $\Lambda$ system. The resonant optical fraction is set by the EOM modulation index $m$ as $2J_1^2(m)$, where $J$ is the Bessel function of the first kind. The laser frequency is stabilized near the center of the two $D_1$ transitions using dual-frequency sub-Doppler spectroscopy~\cite{brazhnikov_dual-frequency_2019}, achieving a fractional frequency stability of $\approx 1\times10^{-12}$ over $10^{4}~\textrm{s}$. The common-mode detuning from the $F^{\prime} = 4$ excited state, $\delta$, is controlled using a pair of acousto-optic modulators.

% Light delivery & optics details:
The atomic beams are interrogated using two parallel laser beams nominally orthogonal to the atomic propagation axis with separation $L \approx 10~\textrm{mm}$. The optical power, $P_{\rm total}$, is split asymmetrically between the two zones, with $\approx 75 \%$ directed to the pump zone to provide strong state preparation followed by a weaker readout pulse~\cite{blanshan_light_2015, micalizio_Raman-Ramsey_2019}. The laser beams are circularly polarized and have $1/e^{2}$ radii $\approx 500 \ (1400)~\mu\textrm{m}$ along (normal to) the atomic beam axis. The device is mounted with its surface normal oriented at $40^{\circ}$ relative to the incident laser beams to minimize light retro-reflected onto the atomic beams. A magnetic field of $\approx 3.6~\textrm{G}$ is applied along the laser wavevector to define the quantization axis and spectrally separate the magnetically-sensitive transitions from the clock transition.

% CPT signal: detection and SNR + clock realization
Typical CPT spectra are shown in Fig.~\ref{fig1}(c) in the case of a narrow laser ($\Gamma_L \ll \Gamma$, black) and a broad laser ($\Gamma_L \approx 9.3~\Gamma$, orange), where half the two-photon (clock) detuning $\Delta =  \omega_{\rm EO} - \omega_0/2$. Here and in the following discussions, all clock frequencies and sensitivities are quoted in the ``half frequency" scale relative to the half-hyperfine, $\sim$ 4.6 GHz EOM modulation frequency. Atomic fluorescence from the second Ramsey zone is detected using a silicon photomultiplier coupled to a 1:1 imaging system with $>10\%$ collection efficiency. The resulting Ramsey fringes exhibit $10$-$15\%$ contrast and a $6~\textrm{kHz}$ full-width at half-maximum (FWHM), set by the Ramsey dark time, $T_{\rm R}$. We realize an atomic clock by locking the EOM modulation frequency to the Ramsey-CPT resonance and comparing the resulting frequency to a local hydrogen maser. The clock exhibits a typical fractional frequency stability at $1~\textrm{s}$ in the low $10^{-10}$ regime when the beam cell is operated near $360~\textrm{K}$, with improved short-term stability achieved through increased atomic flux or improved detection signal-to-noise ratio. 

% Doppler detuning distribution and modification of CPT pumping rates:
The CPT signal and associated clock systematics are strongly influenced by variations in the CPT pumping rate across the atomic ensemble, arising from the Doppler distribution of velocities in the miniature atomic beam. An example of the atomic detuning distribution in the second Ramsey zone is shown by the gray shaded region of Fig.~\ref{fig1}(d), measured using a weak, single-frequency probe beam. The atomic beam signal exhibits a FWHM of $\approx 25~\textrm{MHz}$ set by the chosen imaging geometry. Following \cite{staron_doppler-shift_2026}, the measured beam spectrum is well described by $S(\delta) \propto  e^{-\delta^2/\sigma_{\delta}^2}$, where the Doppler width is given by $\sigma_\delta = k\sigma_v \Delta x / 2L$, with mean squared beam velocity $\sigma_v^2 = 2k_B T/M$. Here, $\Delta x \approx  1.5~\textrm{mm}$ is the spatial extent of the fluorescence collection region along the laser wavevector, $M$ is the mass of $^{133}$Cs, and $k_B$ is Boltzmann’s constant. This Doppler width is significantly larger than the natural linewidth of the $D_1$ transition, in contrast to conventional CSACs where the optical linewidth is typically dominated by GHz-scale collisional broadening.

% Laser linewidth broadening:
To explore the impact of spatially varying pumping rates on clock performance, we controllably broaden the laser spectrum by injecting current noise into the laser diode. The injected noise is spectrally white over $1.5~\textrm{MHz}$ bandwidth, producing Gaussian laser linewidths with FWHM $\Gamma_L$ up to $\approx 10~\Gamma$ (dashed line in Fig.~\ref{fig1}(d)), as measured using the delayed self-heterodyne technique~\cite{okoshi_novel_1980}. The noise bandwidth is sufficiently large that atoms with the most probable transit time across the interaction zones ($\approx $ $2~\mu\textrm{s}$) experience an effectively broadband optical field. Laser broadening over this range makes the CPT pumping rate more homogeneous across the transverse dimension of the atomic beam, and the measured signals and clock stability values are similar to narrow laser interrogation (see Fig.~\ref{fig1}(c)). In addition, laser broadening emulates the spectral profile of a VCSEL (typical linewidth $20$ - $50~\textrm{MHz}$~\cite{gruet_metrological_2013}), which is the laser source envisioned for fully integrated, low-power CSABC implementations.

\section{Light shifts in Ramsey-CPT Spectroscopy}
\label{sec:light_shifts}

Light-based shifts in Ramsey-CPT clocks are different from the linear {\it ac} Stark shifts which appear in many atomic clocks, exhibiting non-linear behavior with laser power, laser detuning, modulation index, and laser linewidth. Here, we summarize the underlying shifts and highlight the impact of spatially-varying CPT pumping rates in the chip-scale beam approach illustrated in Fig.~\ref{fig1}. Broadly, the light-based shifts can be categorized into shifts arising from the ``resonant" frequency components~\cite{hemmer_ac_1989, shahriar_darkstate_1997}, shifts arising from ``off-resonant" interactions~\cite{zhu_theoretical_2000,pollock_ac_2018}, and Doppler-induced microwave phase shifts~\cite{esnault_cold-atom_2013}.  These light-based shifts (here labeled $\Delta_{\rm res}$, $\Delta_{\rm off-res}$, and $\Delta_{D}$, respectively) combine to determine the sensitivity of the clock frequency to the CPT light parameters. We characterize this sensitivity through the coefficients $\chi_\delta = \partial \Delta / \partial \delta$, $\chi_P = \partial \Delta / \partial P$, and $\chi_m = \partial \Delta / \partial m$, which describe the clock response to variations in laser detuning, power, and modulation index, respectively. Managing these sensitivities is critical for realizing long-term clock stability.

\begin{figure}[htbp]
\centering
\includegraphics[width=\linewidth]{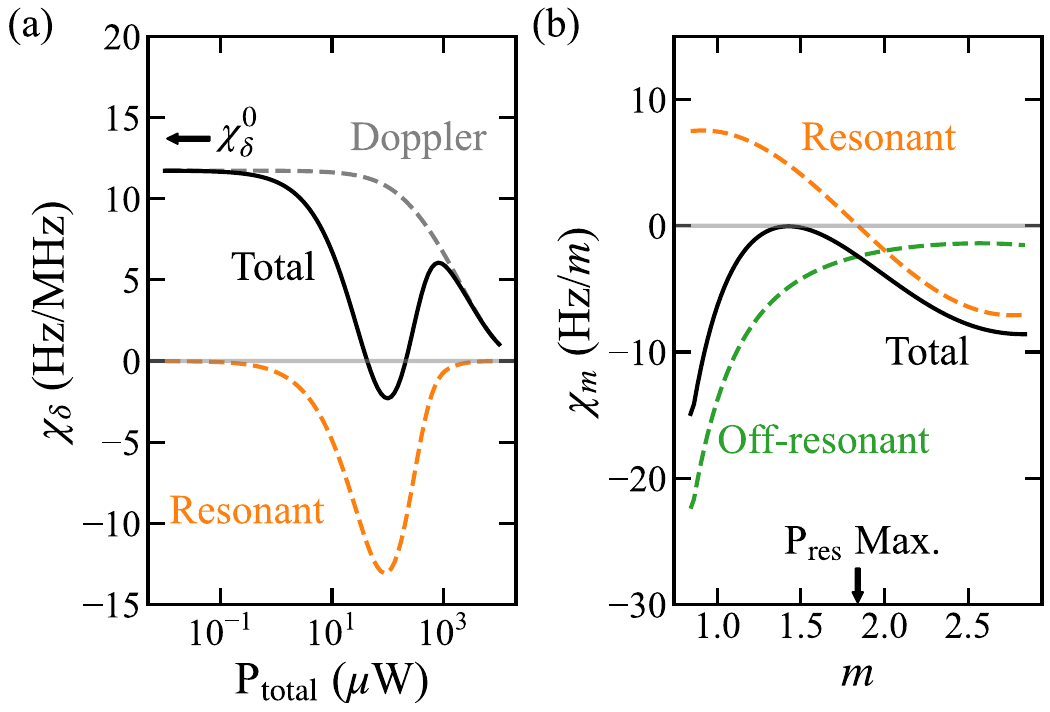}
\caption{Theoretical clock sensitivities to variation in laser frequency (a) and modulation index (b). Resonant shifts (dashed orange), off-resonant shifts (dashed green), and Doppler shifts (dashed gray) combine to realize the total clock sensitivity (solid black), which can be zero under some conditions. The bare Doppler shift ($\chi_\delta^0$) is indicated with an arrow in (a). Finite laser detuning ($\delta \approx \Gamma$) is assumed for the resonant shift in (b).}
\label{fig1.5}
\end{figure}

The pumping dynamics into the CPT dark state determine which atoms in the beam contribute to the clock signal and their associated weights in shifting the clock frequency. Estimated values for the readout probability across the atomic beam are shown in Fig.~\ref{fig1}(d) for our typical experimental parameters, which can be spectrally narrower than the beam distribution at low power, or roughly constant across the beam distribution at high interrogation powers. Following the notation of Ref.~\cite{hemmer_ac_1989}, these estimates are made using the time-varying occupation of the excited state ($F^{\prime} = 4$) in the second Ramsey zone as 
\begin{equation}
\rho_{33} \propto 1- e^{-f\Omega^2\!S \tau_B},
\label{eq:rho33}
\end{equation}
where $\tau_B$ is the transit time in the second Ramsey zone,
\begin{equation}
    \Omega^2\!S
    =
    \frac{\Gamma}{2}
    \frac{\Omega_1^2 + \Omega_2^2}
    {\Gamma^2 + 4\delta^2},
    \label{eq:omega2s}
\end{equation}
is the Raman damping rate that governs dark state formation, and $f = \Gamma/ (\Gamma + 3\Omega^2S)$ is the Raman saturation parameter. Due to the finite collimation of the atomic beam, the transverse velocity distribution produces a spatially varying laser detuning that modifies the Raman damping rate and thus the magnitude of the light shifts. The combination of light-based shifts are discussed below and the key clock sensitivities are illustrated in Fig.~\ref{fig1.5}.

Resonant light shifts are potentially the largest systematic in Ramsey CPT clocks and arise from incomplete pumping into the CPT dark state during the first Ramsey interaction zone~\cite{hemmer_ac_1989}. These shifts can be on the same scale as the Ramsey fringe width and often contain a strong laser-detuning dependence. The resonant shift has been most often studied in the context of ground-state population imbalances in large atomic beams~\cite{hemmer_ac_1989} and cold atom systems~\cite{blanshan_light_2015}, and this shift can be linearly suppressed by reducing the imbalance or exponentially suppressed by increasing the Raman damping rate or pumping time. In a CSABC, no initial population imbalance is expected in the thermal atomic beam, but a population asymmetry can form during CPT pumping due to unequal decay pathways in the CPT $\Lambda$-system~\cite{hemmer_ac_1989}.

The asymmetrical-decay-induced resonant light shift depends on the laser detuning, and this shift has been demonstrated as a convenient method to counteract microwave Doppler shifts and null the laser frequency dependence in a chip-scale Rb beam clock~\cite{staron_doppler-shift_2026}. Assuming initially equal ground-state populations, $\Omega_1^2 = \Omega_2^2$, and operating in the weak-pumping regime ($\Omega^{2}\!S \ll \Gamma$) typical of beam clocks, the resonant phase shift is given by
\begin{equation}
\tan(\phi_{\rm res}) = -r\frac{2\Gamma}{\delta} \
\frac{\sin^{2}(\frac{1}{2}\Omega^2\!S\tau_A\delta/\Gamma)}{e^{\Omega^2\!S\tau_A} - 1},
\label{eq:res}
\end{equation}
where $\tau_A$ is the interaction time in the first Ramsey zone and $r = (\Gamma_1 - \Gamma_2)/(\Gamma_1 + \Gamma_2)$ characterizes the asymmetry in decay rates (see Fig.~\ref{fig1}(b)). The resonant shift is nominally zero at zero laser detuning, and its largest impact on the clock stability is the introduction of a strong laser frequency dependence as $\chi_{\delta}^{\rm res} = \partial\Delta_{\rm res}/\partial \delta$ shown as the orange dashed line in Fig.~\ref{fig1.5}(a), where $\Delta_{\rm res} = \phi_{\rm res}/4\pi T_{\textrm{R}} $, including a factor of 1/2 accounting for half-hyperfine-based frequency notation. This sensitivity is initially zero, builds as scattering generates a population imbalance between the two ground states, and is exponentially suppressed at high Raman damping. 

For interrogation schemes where the sign of the resonant shift (which depends on the sign of $r$) is opposite to the Doppler shift, the two shifts can be balanced to null the overall clock sensitivity to laser frequency $\chi_{\delta}$ as indicated in the solid line in Fig.~\ref{fig1.5}(a). For the Cs $D_1$ line with $\fprime = 4$, we estimate $r \approx 0.3$ and should counteract the Doppler shift, noting that the exact value of $r$ will depend sensitively on the optical pumping dynamics. This resonant shift varies across the atomic beam as the pumping rate and the Doppler detuning change across the sampled atoms, and the impact on the clock was previously shown to be described using a weighted average of Eq.~\ref{eq:res} across the atomic detuning distribution~\cite{staron_doppler-shift_2026}.

Doppler shifts in Ramsey–CPT spectroscopy also introduce a strong laser detuning sensitivity and arise from atomic motion along the CPT $k$-vector, leading to a relative microwave phase shift between the first and second Ramsey zones. This shift is analogous to the ``end-to-end cavity phase shifts” in conventional microwave beam clocks~\cite{mungall_cavity_1972, lee_accuracy_1995}. Our CSABC approach uses co-propagating $\sigma$--$\sigma$ interrogation, and the Doppler shift is given by the ratio of optical to microwave wavelengths ($\chi_{\delta}^{0} \approx 13.7~\mathrm{Hz/MHz}$ or $\approx 3.0 \times 10^{-9}/\mathrm{MHz}$). In practice, the sensitivity $\chi_{\delta}^{\rm D} = \partial \Delta_{\rm D} / \partial \delta$ is somewhat lower than the $\chi_{\delta}^{0}$ and falls off at high probing powers due to spectral broadening of the readout sampling as indicated in the dashed gray line in Fig.~\ref{fig1.5}(a). The Doppler sensitivity introduces stringent requirements for laser frequency stability which are likely infeasible in miniature atomic clocks, and mitigation approaches such as counter-propagating CPT optical fields~\cite{jau_push-pull_2004, esnault_cold-atom_2013} or the previously mentioned shift-balancing using resonant shifts \cite{staron_doppler-shift_2026} are likely required.

Finally, the off-resonant light shifts generate a strong sensitivity to modulation index and arise from the interaction of the modulated optical field with all detuned atomic energy levels, predominantly from the ground and excited-state hyperfine structure. The \textit{ac} Stark shift creates a power-based level shift of the ground state splitting by $\Delta_{ac}$, the magnitude of which changes in a non-linear fashion with the modulation index and the interrogation approach used~\cite{levi_light_2000,zhu_theoretical_2000}. For half-hyperfine interrogation where the two first-order frequency sidebands of the light are resonant with the CPT $\Lambda$-system, this level shift disappears for $m = 2.4$, slightly above the Bessel function maximum at $m = 1.84$.

In Ramsey–CPT spectroscopy, the off-resonant phase shift is set by the phase accumulated from the off-resonant level shift during the two interaction zones as 
\begin{equation}\label{eqn:off_res_shift}
    \phi_{\rm off-res} = A\frac{ \Delta_{ac}}{\Omega^{2}\!S},
\end{equation}
where $A$ is a coefficient that can include contributions from both interaction zones~\cite{pollock_ac_2018}. This ratio is only weakly dependent on the optical power as both $\Delta_{\it ac}$ and $\Omega^2\!S$ scale linearly with the optical power, and the magnitude of the phase shift is controlled by the interrogation scheme and the modulation index. The off-resonant shift has been studied in the context of cold-atom CPT clocks, where $A \approx 1\ \text{--}\ 3$ is predicted for a strong preparation pulse and a weak readout pulse~\cite{pollock_ac_2018}.  Variation of the pumping rate across the sampled atoms can dramatically modify the ensemble-averaged magnitude of  Eq.~\ref{eqn:off_res_shift} as $\Omega^2\!S$ depends strongly on detuning, while $\Delta_{\it ac}$ does not. For simplicity in describing this effect, we absorb the average detuning dependence of $\Omega^2\!S$ into an effective value for A, called $A_{\rm eff}$, which is used with Eq.~\ref{eqn:off_res_shift} assuming $\delta = 0$.
 
For this work, we expect the off-resonant shift to be largely independent of optical power, and potential clock instability to arise primarily from variations in the modulation index. We define a sensitivity to variation in the modulation index as $\chi_{m}^{\rm off-res} = \partial \Delta_{\rm off-res} / \partial m$, where $\Delta_{\rm off-res} = \phi_{\rm off-res}/4\pi T$, again including a factor of 1/2 accounting for half-hyperfine-based frequency notation. For our geometry, we calculate $\chi_m^{\rm off-res} \approx -1.5$~Hz/$m$ ($\approx 3.2\times 10^{-10}/m$) for $A_{\rm eff} = 1$ and $m = 2.4$, implying that control of $m$ is needed at the $10^{-3}$-level. The value of $\chi_m^{\rm off-res}$ is always finite and negative for half-hyperfine modulation, as shown in the green dashed line in Fig.~\ref{fig1.5}(b). Similar to the laser-frequency sensitivity cancellation, the off-resonant modulation index sensitivity can sometimes combine with other $m$-dependent shifts, such as the resonant shift at finite laser detuning (orange dashed line in Fig.~\ref{fig1.5}(b)), to null the overall modulation index sensitivity.

\section{Results and Discussion}
\label{sec:results_and_discussion}

\begin{figure}[htbp]
\centering
\includegraphics[width=\linewidth]{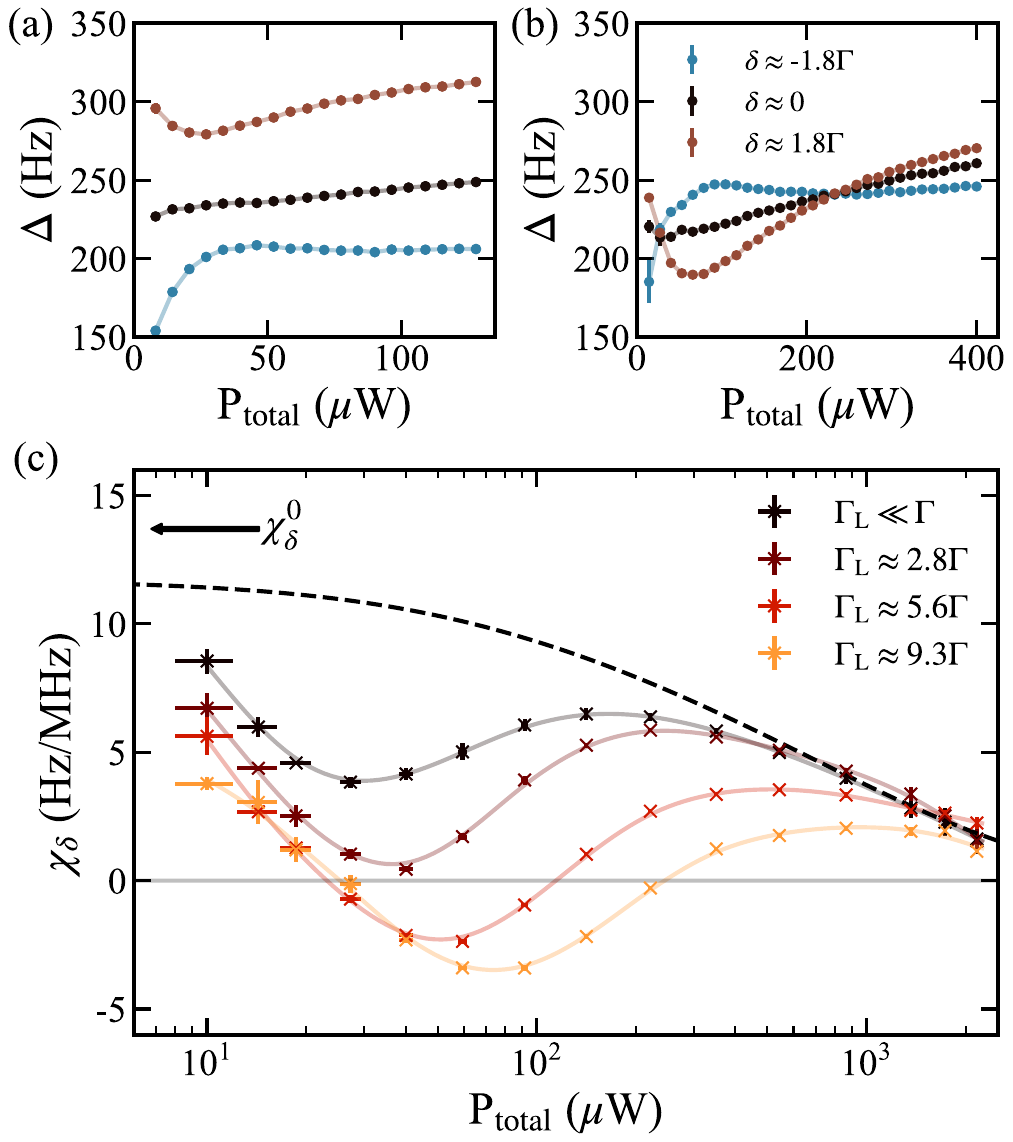}
\caption{Doppler shifts and resonant light shifts. The clock shift vs. laser power is measured at three common-mode laser detunings using narrow (a) and broad (b) laser sources. (c) The clock's laser detuning sensitivity $\chi_{\delta}$ is plotted as the laser spectrum is broadened. The black and yellow clock data correspond to (a) and (b), respectively. The bare Doppler shift sensitivity $\chi_\delta^0$ (arrow) and a numerically estimated narrow-laser limit to the Doppler shift (dashed line) are included for comparison. Error bars indicate standard uncertainty, some are smaller than the marker. Solid lines are guides to the eye.}
\label{fig2}
\end{figure}

% Fig. 2 (a) and (b): $\Delta$ vs. $P_{\textrm{total}}$ at various $\delta$
We  experimentally probe the light-based systematics in our CSABC, starting with the combination of resonant and Doppler shifts shown in Fig.~\ref{fig2} using spectrally narrow (a, $\Gamma_{\rm L} \ll \Gamma$) and spectrally broad (b, $\Gamma_{\rm L} \approx  9.3\Gamma$) laser excitation. The clock frequency is plotted as a function of total optical power for three values of $\delta$ for these two cases, revealing a strong linear dependence on the laser detuning. Non-linear, power-dependent shifts of the clock frequency are observed at finite detuning arising from resonant shifts induced by asymmetrical decay described in Eq.~\ref{eq:res}. The narrow-linewidth data is dominated by Doppler shifts, which induces a large sensitivity to laser detuning. The broad-linewidth data shows similar apparent magnitudes for the Doppler and resonant shifts, leading to cancellation of the laser sensitivity at certain optical powers (near 20 and 220 $\mu$W in Fig.~\ref{fig2}(b)), similar to our past demonstration of a Doppler-mitigation strategy in a Rb-based CSABC~\cite{staron_doppler-shift_2026}. We note that the clock frequency is centered around 240 Hz offset from the expected transition frequency, which we ascribe to a small asymmetry in the beam splitter path lengths. 

To explore this change in sensitivity to laser frequency deviation $(\chi_\delta)$ with laser broadening, we measure $\chi_{\delta}$ as a function of laser power for a range of laser linewidths as shown in Fig.~\ref{fig2}(c). Here, $\chi_{\delta}$ is calculated using clock measurements spanning 8~MHz around $\delta = 0$, where the shifts are well described by a linear model. The observed sensitivities are less than expected Doppler shift $(\chi_{\delta}^{0})$ and consists of a competition between Doppler shifts and resonant light shifts induced by asymmetrical decay which vary with optical power as illustrated in Fig.~\ref{fig1.5}(a).  Broadening the laser reduces the effective value for $\Omega^2S$ across the sampled atoms and pushes the minimum in $\chi_{\delta}$ to higher powers. This is seen in Fig.~\ref{fig2}(c), where the minimum of $\chi_{\delta}$ (maximum negative value for $\chi_{\delta}^{\rm res}$) occurs near 25~$\mu$W  for the narrow laser data (black curve), while the broadest laser measured has a minimum for $\chi_{\delta}$ near 80 $\mu$W (orange curve). 

The largest effect in Fig.~\ref{fig2}(c) is a strong modification of the Doppler shift that causes $\chi_{\delta}$ to progressively decrease as the laser linewidth is broadened. The Doppler shift is set by the average microwave phase shift between the two Ramsey zones, which in our experiment is the average relative position (or effective detuning) of the atom probed in the two zones. The possible variation in the average detuning can be understood by considering the spatial sampling of atoms within the atomic beam shown in Fig.~\ref{fig1}(d), where broadening the laser (dashed purple line) or increasing the optical power (green line) makes the pumping more homogeneous across the atoms relative to the narrow, low-power case (solid purple line) and reduces the possible Doppler shift.

We numerically estimate the Doppler shift in the presented clock geometry as 
\begin{equation}
\chi^{\rm D}_{\delta} = \chi_{\delta}^{0}\frac{\partial\left<  \delta \right>}{\partial \delta},
\label{eq:doppler}
\end{equation}
where the brackets indicate a weighted average over the experimentally measured beam profile $S(\delta)$ (shaded gray region in Fig.~\ref{fig1}(d)) and the theoretical saturated readout probabilities set by Eq.~\ref{eq:rho33}. For narrow laser excitation, the effective Doppler shift from Eq.~\ref{eq:doppler} is shown as the dashed line in Fig.~\ref{fig2}(c). Here, the Doppler shift is less than $\chi_{\delta}^{0}$  at low $P_{\rm total}$ due to the finite widths of $S(\delta)$ and $\Gamma$. The Doppler sensitivity decays at high $P_{\rm total}$ as the readout probability saturates across the atomic beam, reducing the effect of variations in the averaged sampled laser frequency. The numerical estimate for $\chi^{\rm D}_{\delta}$ agrees with the measured, narrow laser value for $\chi_{\delta}$ (black points in Fig.~\ref{fig2}(c)) at high power with no free parameters, indicating that the model captured the essential readout saturation behavior. This model in Eq.~\ref{eq:doppler} can be extended to account for broadened laser spectra, and an estimate for $\chi_{\delta}^{\rm D} \approx 4.0$~Hz/MHz is found for the $\Gamma_L \approx 9.3 \Gamma$ data (orange curve), consistent with the measured data. 

The detuning sensitivities at intermediate values of laser linewidth (maroon and red lines) are also plotted in Fig.~\ref{fig2}(c) and show similar magnitude for the resonant shift ($\approx 7$~Hz/MHz) and intermediate reductions of the Doppler shifts, determined by the laser spectral width. The magnitude of the resonant shift ``dip" in $\chi_{\delta}$ is largely independent of the laser broadening as the shift magnitude is primarily set by $r$, which is a property of the atom and the interrogation scheme used. The effect of broadening the laser on the resonant shifts is simply to shift the dip minimum to higher laser power due to the broadening-induced decrease in the Raman damping rate. Here, the compensation of the Doppler shift with the resonant light shift to enable $\chi_{\delta} = \chi^{\rm D}_{\delta} + \chi^{\rm res}_{\delta}$ to pass through zero can be tuned with the laser linewidth (or equivalently by varying the atomic beam properties), and the dependence of $\chi_{\delta}$ on laser power can be reduced if desired by engineering the dip in $\chi_{\delta} $ to be near zero, as shown in the maroon data in Fig.~\ref{fig2}(c). Similar to Rb, the $\chi_{\delta} = 0$ points occur conveniently for the higher hyperfine ground state, where the Clebsch-Gordan coefficients and signal contrast are larger than for the lower excited state. Measurements using $F' = 3$ show enhanced laser detuning sensitivity, as $r$ is negative for this state and shifts in the same direction as the Doppler shift.

In addition to the Doppler-mitigation strategy, the combination of the Doppler shifts and resonant light shifts can also be used to realize ``doubly insensitive" operation where the sensitivity to laser detuning is also nulled by operating at finite common-mode detuning. This cancellation uses the power-dependence of the resonant shift to cancel other power-dependent shifts, and the sign of this cancellation is controlled by the sign of the laser detuning. An example of this ``doubly insensitive" operation is shown in the blue curve in Fig.~\ref{fig2}(b) near $P_{\rm total} \approx 220 \ \mu$W, where the power sensitivity $\chi_{P} < 30$~mHz/$\mu$W and $\chi_{\delta} < \ 50$~mHz/MHz. We note that the expected value for $\chi_P$ is zero at $\delta = 0$ for all of the light-based shifts (see Eq.~(\ref{eq:res}) and Eq.~\ref{eqn:off_res_shift}), and doubly insensitive operation should then be expected at $\delta = 0$. Instead, we observe a finite power shift of $\sim 0.1 \ {\rm Hz}/\mu$W (see black lines in Fig.~\ref{fig2}(a,b)) which we attribute to a small misalignment of the fluorescence imaging optics that induces a slightly asymmetric sampling of the Doppler shift.

\begin{figure}[htbp]
\centering
\includegraphics[width=\linewidth]{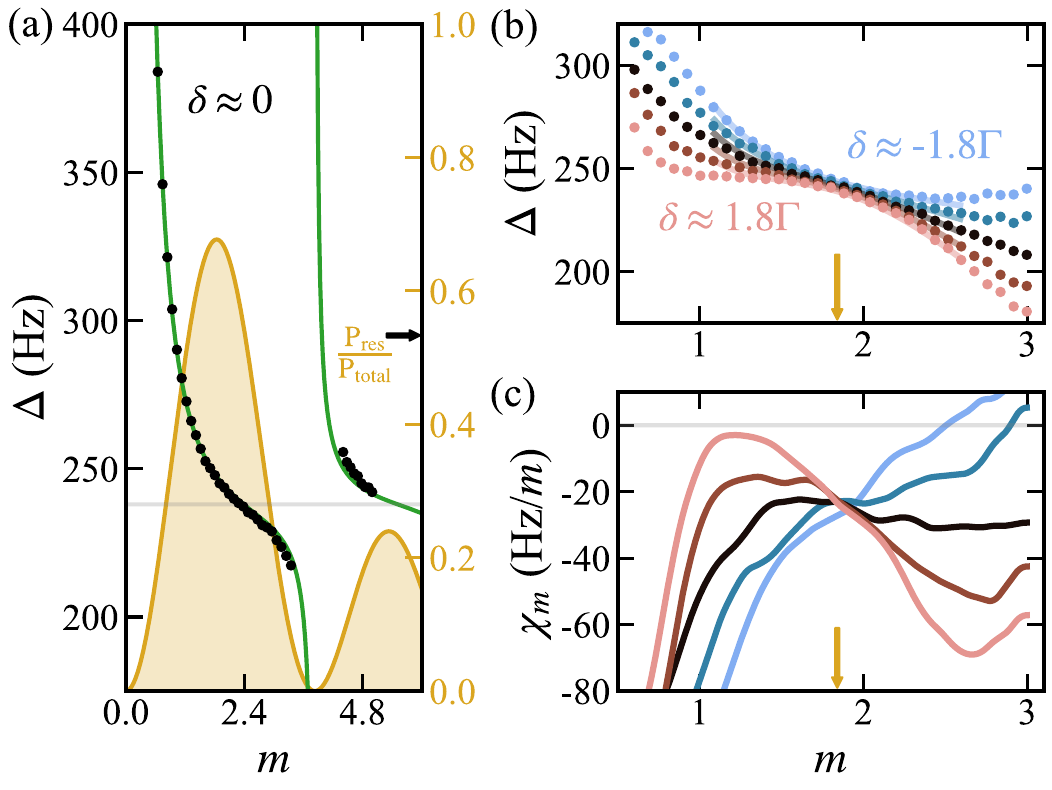}
\caption{Off-resonant and resonant light shifts. (a) The clock shift at fixed resonant optical power (black points) matches a scaled version of the off-resonant shift (green line, Eq.~\ref{eqn:off_res_shift} with $A_{\rm eff} = 9.4$). The shaded area indicates the fractional optical power in the resonant sidebands. (b) The clock shift at fixed $P_{\rm total}$ shows a mixture of resonant and off-resonant shifts vs. modulation index, which are well described by Eq.~\ref{eqn:combined_shift} (solid lines). (c) The sensitivity to modulation index can vary with laser detuning due to the cross-talk between shifts, which disappears at the peak of the resonant power (gold arrows). 
}
\label{fig:offres}
\end{figure}

The remaining light-based systematic is the off-resonant light shift, which we measure near the $P_{\rm total} \approx 220 \ \mu$W doubly-insensitive point identified using broad-laser interrogation. The off-resonant shift is predominantly power-independent as shown in Eq.~\ref{eqn:off_res_shift} and controlled by the modulation index of the light. Variation in $m$ at constant $P_{\rm total}$ impacts the resonant power and the magnitude of the resonant light shifts, which creates cross-talk between $m$ and the power-based shifts. We isolate the off-resonant shifts by adjusting $P_{\rm total}$ to keep the resonant optical power $P_{\rm res} = 2J_1^2(m) P_{\rm total}$ (and hence $\Delta_{\rm res}$) constant as the clock frequency is recorded at varying $m$ as shown in Fig.~\ref{fig:offres}(a). The resonant power stability and the value of $m$ are calibrated using a scanning Fabry-Perot cavity for this data, and $P_{\rm res}$ is constant across the measured range of $m$ within $1\%$.

The isolated off-resonant shift in Fig.~\ref{fig:offres}(a) matches well with the theoretical shift in Eq.~\ref{eqn:off_res_shift} with $A_{\rm eff} \ = 9.4$, including the $\approx 240 \ {\rm Hz}$ frequency offset from the beam splitter asymmetry. The large value for $A_{\rm eff}$ relative to cold-atom experiments where the measured $A \approx 2$~\cite{pollock_ac_2018} is explained by the reduced values for $\Omega^{2}\!S$ in Eq.~\ref{eqn:off_res_shift} due to laser broadening and the sampled distribution of atomic detunings. This larger off-resonant shift introduces a significant clock sensitivity to modulation index $\chi_m  \approx -13.5$~Hz/$m$ at $m = 2.4$, requiring a modulation index stability of $3\times10^{-4}$ to reach $10^{-12}$ fractional clock stability.

In a fully-integrated CSABC, $P_{\rm total}$ and not $P_{\rm res}$ will be stabilized, and a mixture of off-resonant and resonant light shifts will be observed at varying $m$. This combination of shifts at fixed $P_{\rm total}$ is shown in Fig.~\ref{fig:offres}(b) at several laser detunings. The data show similar behavior to the isolated off-resonant light shifts in Fig.~\ref{fig:offres}(a), with detuning-dependence introduced by the resonant light shifts. The data around $m = 1.84$ are well described by a sum of the two shifts using a linearized model for the resonant light shifts as 
\begin{equation}\label{eqn:combined_shift}
    \Delta = \Delta_{\rm off-res}(m) + \chi_P(\delta) P_{\rm res}(m), 
    \end{equation}
where $\chi_P(\delta)$ is the local, measured resonant power sensitivity. This model is plotted as solid lines in Fig.~\ref{fig:offres}(b) using $\chi_P$ values measured at $m \approx 1.84$ and agrees well with the measured clock frequencies within a restricted range of $m$ where $\Delta_{\rm res}$ is linear. The data converge near $m = 1.84$ (the peak of the $P_{\rm res}$), where the variation of optical power and the resonant light shift are locally-insensitive to $m$.

The clock sensitivity to $m$ is shown in Fig.~\ref{fig:offres}(c), and the combination of light shifts generates passively stable operation points at some non-zero laser detunings where $\chi_m = 0$. ``Triply insensitive" operation points with simultaneous zero-crossings of $\chi_m $, $\chi_{\rm P}$, and $\chi_\delta$ have been found using this approach, but exploiting them requires cross-talk between the resonant and off-resonant shifts which complicates optical power and modulation index stabilization and is likely not desirable for low-drift clock operation.  CPT clocks often operate at the zero point for the off-resonant shift (here $m = 2.4$), but given the relative power-insensitivity of the off-resonant shift in our approach, it is advantageous to operate at the peak of the Bessel function ($m = 1.84$) where the cross-talk between the modulation index and $P_{\rm res}$ is minimal. This moderately increases $\chi_m $ to $\approx -23$~Hz/$m$ and maximizes $P_{\rm res}$.

\begin{figure}[htbp]
\centering
\includegraphics[width=\linewidth]{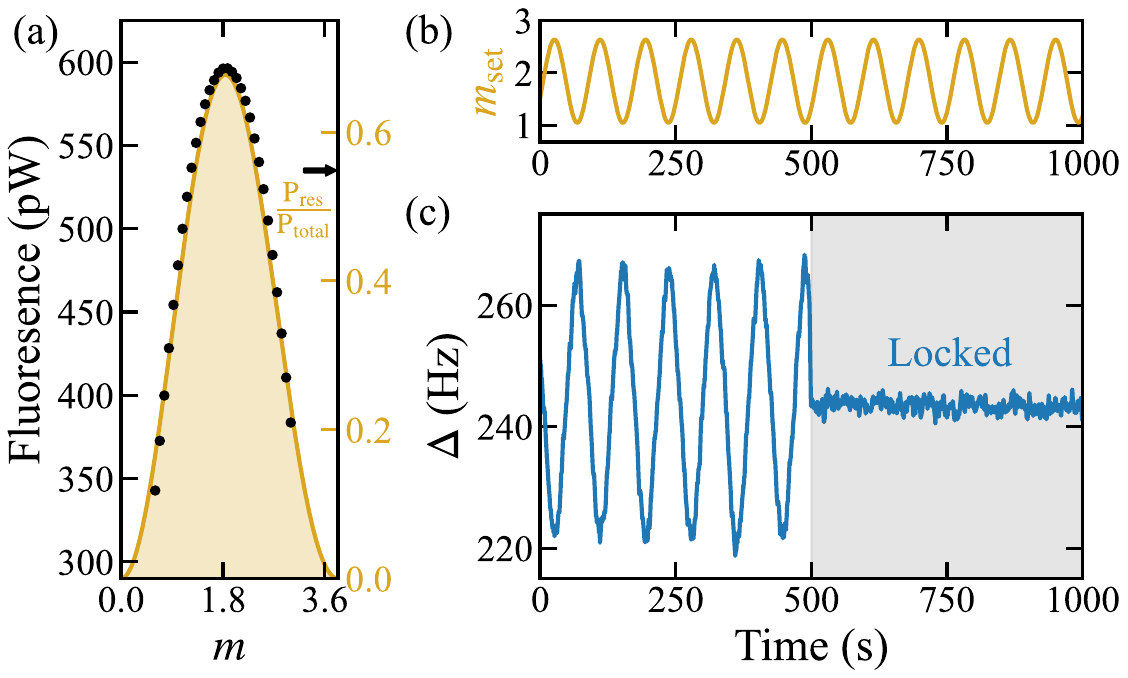}
\caption{Active modulation index stabilization. (a) Atomic beam fluorescence measured away from the CPT resonances provides a faithful measurement of $m$. (b) Sinusoidal modulation of the modulation index setpoint $m_{\rm set}$ leads to a time-varying clock frequency (c). This variation of $m$ is suppressed when active feedback is applied (shaded area).}
\label{fig:mod}
\end{figure}

Considering the measured light shifts holistically, the operational mode which is most promising to achieve long-term stability is the broad laser ``doubly insensitive" operation point with near-zero sensitivity to laser power and laser detuning at $m = 1.84$. Stringent control of $m$ is required at this point and can be accomplished using one-photon fluorescence spectroscopy of the atomic beam as an {\it in situ} probe of $m$ as shown in Fig.~\ref{fig:mod}(a). Here, we adjust $\omega_{EO}$ away from the CPT resonances at $\Delta \approx 500~\textrm{kHz}$, where the measured fluorescence scales with $P_{\rm res}$ and faithfully reproduces its Bessel function dependence on $m$. Utilizing lock-in detection of the signal maximum, the $m$ discriminator slope is estimated $> 300~\textrm{pW}/m$. For a $1\%$ duty cycle in this approach, $m$ stability can be better than $10^{-3}$ at 1 s, which is sufficient to compensate for relatively large values of $\chi_m$.

To demonstrate active stabilization of $m$ using this lock-in approach, we applied a sinusoidal modulation to the $m$ setpoint $m_{\rm set}$ as shown in Fig.~\ref{fig:mod}(b). For this test, $m_{\rm set}$ was varied by $\pm 0.8$ around $m = 1.84$ and the clock moved by $\approx \pm 20 $~Hz without feedback applied to $m$ as shown in Fig.~\ref{fig:mod}(c), consistent with the measured value for $\chi_m$ in Fig.~\ref{fig:offres}(c). After 500 s, we used a digital feedback signal generated from the lock-in detected peak fluorescence summed with $m_{set}$ to control $m$. The clock signal for ``locked" operation is shown in the shaded area of Fig.~\ref{fig:mod}(c). Here, the clock modulation was reduced by $\approx 50$ times relative to the unlocked case, limited by a time delay in the feedback. Rejection will likely be significantly higher in typical clock operation, where only long-term drift of $m$ is important. Lock-in stabilization of $m$ requires the ability to actively control RF power and introduces a small amount of dead-time to the clock, as implemented here.

\begin{figure}[htbp]
\centering
\includegraphics[width=\linewidth]{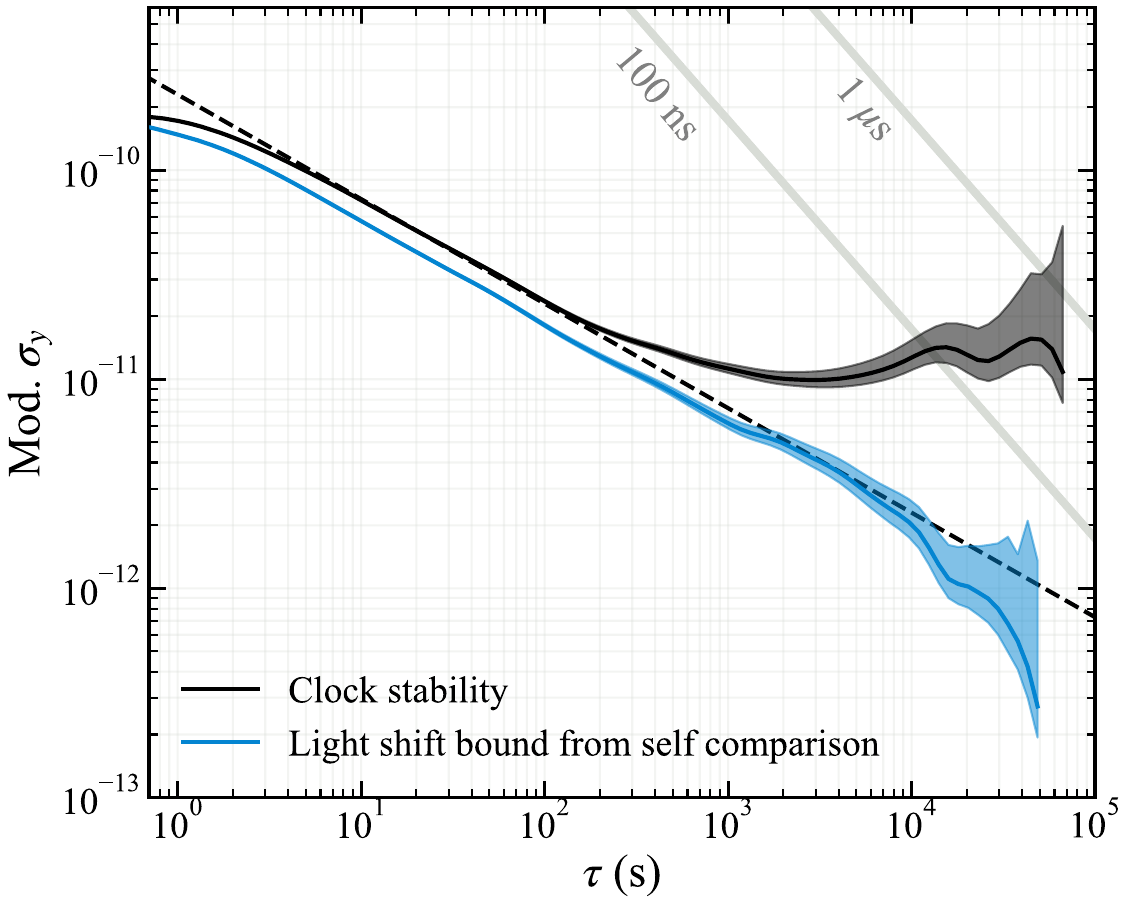}
\caption{Clock stability (black) using doubly-insensitive operation point. Shaded band indicates stability uncertainty. Dashed line indicates ${\rm Mod} \ \sigma_y = 2.2\times10^{-10}/\sqrt{\tau}$. The modulation index-based systematic (which dominates the light-based shifts) is inferred using interleaved self-comparison (blue). }
\label{fig5}
\end{figure}

Utilizing the ``doubly insensitive" operation point with a broad laser ($\Gamma_{\rm L} \approx 9.3 \Gamma$), we measure the CSABC stability over several days as shown in the black line in Fig.~\ref{fig5}. The fractional frequency stability is characterized using the modified Allan deviation and the short-term stability is mod $\sigma_y$ $\approx 2.2\times10^{-10}$ at averaging time $\tau = 1~\textrm{s}$. The clock averages down for about $10^3~\textrm{s}$ before hitting an apparent flicker floor at $\approx 1\times10^{-11}$, where it remains out to $\approx 6\times10^{4}~\textrm{s}$. The floor appears coupled to the clock's thermal environment with a sensitivity of about $12~\textrm{Hz}/\textrm{K}$, likely an artifact of our lab-scale apparatus that could be surpassed with further device integration or refined thermal management. We have characterized the dominant physics-based systematics through out-of-loop measurements of laser frequency, power, modulation index, and magnetic field fluctuations, demonstrating control of the associated clock frequency shifts below the $10^{-12}$ level over relevant timescales. 

We have also evaluated the key light-based instabilities using interleaved self-comparison to isolate specific systematics and provide direct, atom-based measurements of their impact on clock stability. For these comparisons, two independent clock signals are formed at operating points with significantly different sensitivities to laser frequency, power, or modulation index while other effects, including thermal drift, remained largely common mode. The two clock signals are recorded in an interleaved fashion (here at $\approx 4 $~Hz) and the difference frequency isolates drift arising from the isolated parameter. 

We show an example of interleaved self-comparison isolating the modulation index sensitivity as the blue line in Fig.~\ref{fig5}. Here, the interleaved clock signals were recorded at $\delta = -2.2~\Gamma$ with $m = 1.18$ ($\chi_m ^{\rm high}\approx -65$~Hz/$m$) and $m = 2.5$ ($\chi_m^{\rm low} \approx 3$~Hz/$m$), similar to the blue data in Fig.~\ref{fig:offres}(c). The difference of these clock frequencies provides a measure of the modulation index, and the inferred modulation index drift is converted to a clock instability limit by scaling the difference frequency stability by $\chi_m^{\rm clock}/(\chi_{m}^{\rm high} - \chi_{m}^{\rm low}) \approx 0.34$, where $\chi_m ^{\rm clock}\approx -23 \ {\rm Hz}/m$ is the sensitivity where the clock is operated. The isolated $m$-based systematic is characterized using the modified Allan deviation and plotted as the blue line in Fig.~\ref{fig5}. The inferred clock instability shows white noise-limited averaging out to $5\times10^4$ s and provides an upper bound for the leading light-based systematic in the low $10^{-12}$ range, limited by the sensitivity of this approach.

Clock self-comparisons for laser detuning and laser power sensitivities were similarly performed using operating modes at two $P_{\rm total}$ values with large differences in $\chi_{\delta}$ and $\chi_P$, respectively. White noise-limited averaging beyond $10^4$ s was observed for both measurements, similar to the results for the $\chi_m$ self-comparison, with inferred clock instability limits from these systematics near the $10^{-13}$ level, set by the low operational values for $\chi_{\delta}$ ($< 50~\textrm{mHz/MHz}$) and $\chi_P$ ($< 30~\textrm{mHz}/\mu\textrm{W}$). 

We have also used the lock-in detection scheme for stabilizing $m$ presented in Fig.~\ref{fig:mod} as an additional atom-based witness for the modulation index. This technique has significantly higher signal-to-noise ratio than the self-comparison tests and was performed without active feedback to the modulation index, using the lock-in error signal to infer the value for $m$. This test shows the modulation index is stable at the $5\times10^{-5}$ level from 100 s to $5\times10^4$ s, with an inferred clock instability well below the $10^{-12}$ level over the same timescale. These results together demonstrate control of the light-based systematics at or below the $10^{-12}$ level and show promise for long-term stable CSABC operation. 

\section{Conclusions and Outlook}
\label{sec:conclusions_and_outlook}

This work demonstrates the combination of resonant shifts, off-resonant shifts, and Doppler shifts that exist in a CSABC and shows how these shifts can be exploited to substantially reduce or manage key clock sensitivities without increasing device complexity. The role of varying pumping rate across the atomic beam is shown to exaggerate both the resonant and off-resonant light shifts by decreasing the average Raman damping rate across the atomic beam, and these shifts are well captured through simple modeling utilizing the spectral width of the atomic beam and the interrogating laser. We further show that operation near doubly-insensitive points, together with active modulation index stabilization based on atomic beam spectroscopy, offers a straightforward path for managing the dominant clock systematics and realizes sub-$\mu$s stability for day-scale holdover, even in a large and relatively thermally unstable laboratory demonstration. 

These results suggest several broader considerations for future CSABC implementations. First, the shift balancing which enables operation with nulled laser detuning and power sensitivity appears broadly achievable, as the magnitude of the Doppler and resonant light shifts can be tuned using properties of the atomic beam and the laser interrogation.  Second, maintaining the optical power at the doubly-insensitive point at long-integration times is important, and this differs from the optical power stability requirement in many CPT-based clocks with linear light shifts\cite{yudin_combined_2018, shuker_reduction_2019, yudin_general_2020, shuker_ramsey_2019, zhu_theoretical_2000, shah_continuous_2006, mejri_atomic_2016, zhang_rubidium_2016}. Here, the non-linear shift with laser detuning suggests that $\chi_{\delta}$ could be actively measured during clock operation and used to stabilize the clock at this operation point using a slow feedback loop, potentially removing long-term drift of laser power similar to schemes that are used to stabilize the modulation index in CSACs~\cite{zhu_theoretical_2000}. Finally, as the Doppler shift can be strongly suppressed at high laser broadening, operation with a high laser linewidth and a more tightly collimated atomic beam could be a convenient alternative for strongly attenuating the detuning sensitivity generated from resonant and Doppler shifts. This approach would likely also reduce the optical power sensitivity, but may require higher laser power~\cite{ii_high-power_2022, mu_53_2025} and impact the clock's short-term stability due to a lower expected signal level. 

Ongoing work is focused on transitioning CSABCs into integrated systems which overcome the thermal limitations of the current implementation and demonstrate low-SWaP operation. Approaches are being explored to enable further miniaturization and wafer-scale fabrication of atomic beam devices~\cite{kelleher_wafer-scale_2026}, as well as microwave oscillators and control circuits which support this effort are being developed~\cite{russell_low_2025, russell_rf_2025, russell_rf_2026}. More broadly, several promising alternatives are also being pursued by other groups toward the same low-SWaP timing stability goal, including improved CSAC operation \cite{abdel_hafiz_symmetric_2018, carle_pulsed-cpt_2023,rivera-aguilar_microcell_2025}, miniaturized laser atomic oscillators \cite{pandey_cesium_2026}, and Lam\`e-mode MEMS oscillators~\cite{yang_precision_2025, yan_micromechanical_2026, xu_frequency_2026}. Miniaturized atomic beams are also being explored for applications beyond timing, with recent demonstrations of laser cooling experiments~\cite{li_stimulated_2023} and non-classical light generation based on cavity QED systems~\cite{larsen_chip-scale_2026}. Future applications in Rydberg sensing and atom interferometry may further expand the utility of chip-scale atomic beams as a versatile, low-SWaP architecture for precision measurement and quantum sensing.

% \section{}
{\it Acknowledgments:} This work was supported by NIST and the Defense Advanced Research Projects Agency (DARPA) H6 program. AS and ML acknowledge financial assistance award 70NANB18H006 from the U.S. Department of Commerce, NIST. The views, opinions and/or findings expressed are those of the authors and should not be interpreted as representing the official views or policies of DARPA or the U.S. Government. Distribution Statement ``A" (Approved for Public Release, Distribution Unlimited). This document has not been peer reviewed but has been cleared by NIST for release.

{\it Data availability:} The data are available from the authors upon reasonable request.

\bibliography{manual.bib}

\end{document}